\documentclass[%
 reprint,
 amsmath,amssymb,
 aps,
]{revtex4-2}

\usepackage{graphicx}
\usepackage{dcolumn}
\usepackage{bm}

\usepackage{color}
\usepackage{float}
\usepackage{amsmath}
\usepackage{stackrel}
\usepackage{graphicx}
\begin{document}

\preprint{APS/123-QED}

\title{Unpredictable and Uniform RNG based on time of arrival
using InGaAs Detectors}

\author{Anindita Banerjee}%
 \email{anindita@qunulabs.in}
\affiliation{%
 QuNu Labs Pvt Ltd., M.G. Road, Bangalore, India
}
\author{Deepika Aggarwal}%
\affiliation{%
 QuNu Labs Pvt Ltd., M.G. Road, Bangalore, India
}
\author{Ankush Sharma}%
\altaffiliation[Also at ]{%
BITS Pilani,  K. K. Birla Goa Campus, Zuarinagar, Goa, India
}

\author{Ganesh Yadav}%

\affiliation{%
 QuNu Labs Pvt Ltd., M.G. Road, Bangalore, India
}
\date{\today}
\begin{abstract}
Quantum random number generators are becoming mandatory in a demanding technology world of high performing learning algorithms and security guidelines. Our implementation based on principles of quantum mechanics enable us to achieve the required randomness. We have
generated high-quality quantum random numbers from a weak coherent
source at telecommunication wavelength. The entropy is based on time of arrival of quantum states  within a predefined time interval. The detection of photons by the InGaAs single-photon detectors and  high precision time measurement of 5 ps
enables us to generate 16 random bits per arrival time which is the highest reported
to date. We have presented the theoretical analysis and experimental
verification of the random number generation methodology. The method eliminates the
requirement of any randomness extractor to be applied thereby, leveraging the principles of quantum physics
to generate  random numbers. The output data rate is on an average of
2.4 Mbps. The raw quantum random numbers are compared with  NIST prescribed Blum-Blum-Shub pseudo random
number generator and an in-house built hardware random number generator
from FPGA, on the ENT and NIST Platform.
\end{abstract}
\maketitle


\section{Introduction:\label{Introduction: }}

The importance of  random numbers was realised  much early in human society,
its requirement was governed by the beliefs and socioeconomic structure.
The methods of generating random numbers evolved with the development
of science and technology. In the present era, random numbers play
a significant role in  statistical analysis, stochastic simulations, cyber security applications, gaming, cryptography, and many others. Random numbers can be generated by two approaches,  a software approach  termed as  pseudo random number generator (PRNG)  is based on mathematical algorithm and a hardware approach  termed as true random number generator (TRNG) that can extract randomness from physical processes. In software approach, we cannot deny the possibility of backdoor, reapplication of seed generates same random numbers repeatedly and a weak entropy can substantially compromise the security system. These issues can be overcome by  TRNG. However, if the physical process is  classical in nature it employs  causality behind complexity.  A TRNG based on quantum physics is termed as quantum random
number generator (QRNG). The probabilistic nature of quantum mechanics can be employed to generate true random
numbers that are unpredictable, irreproducible, and unbiased. It generates random numbers as a result of  measurement
on a quantum system. The quality of quantum random numbers is strongly
dependent on the properties or behavior of the quantum entity and the elimination of classical noise.

A random bit sequence is characterized
by two fundamental properties i.e. uniformity and unpredictability,
of which the latter is the most important. Uniformity is achievable
by mathematical algorithms, however, for unpredictability, none other
than the inherent randomness of quantum mechanics can be trusted.   Quantum random numbers can be  generated from  several  sources, for example, radioactive decay
\cite{Schmidt}, the quantum mechanical noise in electronic circuits
known as shot noise \cite{Shen}, measuring and digitizing photon
arrival times \cite{Wayne,stipcevic}, quantum vacuum fluctuations \cite{Gabriel},
laser phase fluctuations \cite{Guo}, optical parametric oscillators
\cite{Marandi}, amplified spontaneous emission \cite{Williams} etc.
Several optical QRNG  schemes  \cite{Herrero} have been proposed on the principle of  time of arrival of photon. The  arrival time of photon is considered as a quantum random variable and it  can generate $n$ random bits where,  $n$ depends on the precision of time measurement. Software \cite{Wayne,wahl,discretenco}  and hardware \cite{Wayne2}  approaches  were investigated to minimise (eliminate) the bias and improve the quality of throughput from the time of arrival entropy. The authors in \cite{Nie} showed that when  an external time reference is used, the raw random numbers are generated from the photon arrival in time bins within the external time reference and they are uniformly distributed in time. Hence,  we can consider this quantum entropy source to be one of the ideal candidates for TRNG.

In this paper,
we have reported  our work on QRNG based on time of arrival (ToA) principle using an external time reference. We have implemented the scheme using InGaAs detectors. We have used a different method for generating random numbers and could extract $16$ random bits per detection event. This is the highest recorded entropy per detetion. In our work, we have adhered to the quantum noise random number generator
architecture  recommended by ITU-T X.1702 \cite{ITU} and
this is presented in Fig. \ref{fig:QRNG-architecture}. The raw
data is extracted by performing a measurement on a quantum state which
can be a photon and we are deriving the random numbers from the data acquisition
process. The quantum entropy source (QES) in the presented work belongs
to subclass 1 where minimum entropy is accessed by estimating the
implementation imperfection. The quantum state is prepared using an
optical process and the quantum measurement is based on the Poisson nature
of photon detection by single-photon detectors (SPD). Raw data is
generated by digitizing the output from the single-photon detector. Continuous
monitoring of laser parameters, detector parameters, and amplitude
of the quantum signal at the detector enables assessment of entropy for
evaluation of quantum randomness in the random number sequence. The
implementation imperfections lead to an increase in classical noise therefore, these are identified, continuously monitored, and eliminated.

In section \ref{sec:Source-of-randomness}  we have explained the source
of quantum randomness, in section \ref{sec:Time-of-arrival} we have
discussed the principle of time of arrival in detail along with its
theoretical analysis and sources of bias in the implementation. In
section \ref{sec:Experimental-analysis} we have presented the experimental
setup, in section \ref{sec:Post-Processing} we have discussed the
entropy estimation and we have concluded in section \ref{sec:Conclusion}.

\begin{widetext}
\begin{center}
\begin{figure}[b]

\includegraphics[scale=0.3]{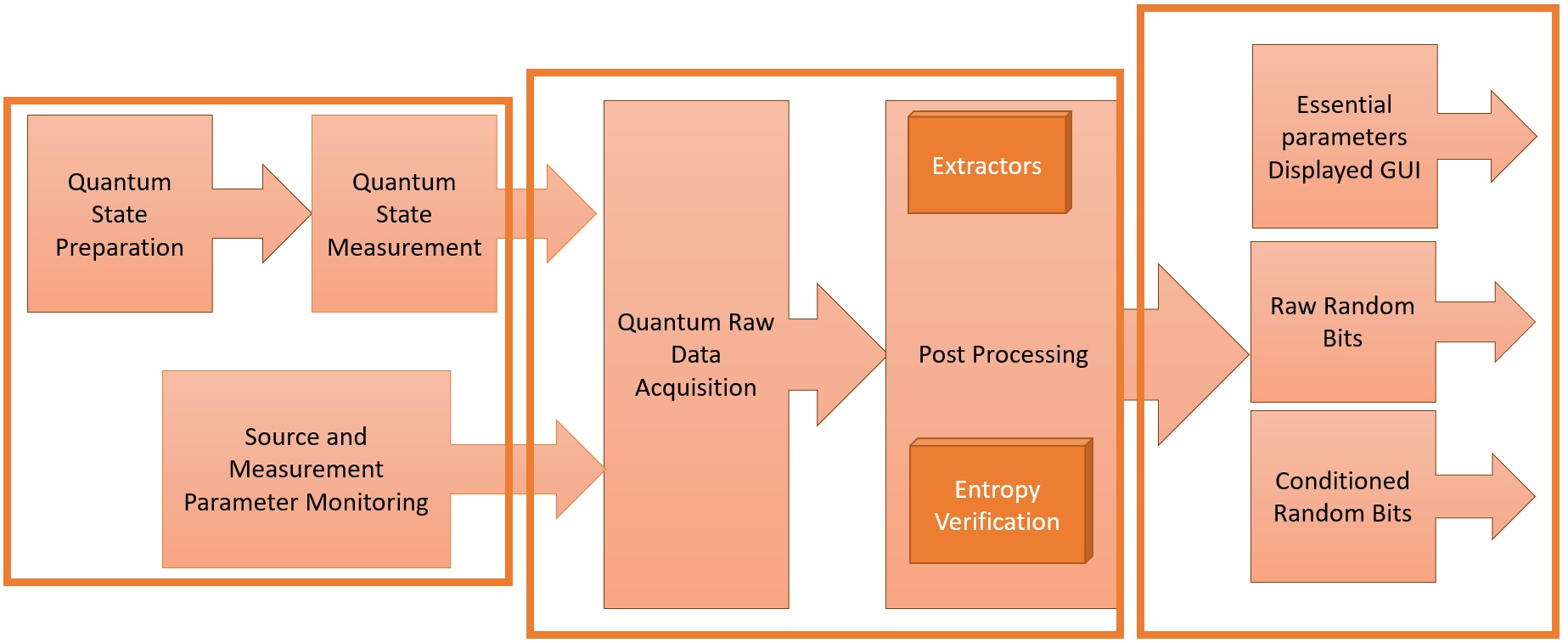}
\caption{\label{fig:QRNG-architecture} Quantum Noise Random Number Generator \cite{ITU}.}

\end{figure}
\end{center}
\end{widetext}
\section{Source of randomness \label{sec:Source-of-randomness}}

The quantum randomness in QRNG comes from the consumption of coherence.
Since our QRNG method is based on the time of arrival principle hence,
it is important to briefly discuss the coherent state and present a mathematical
description of the photon. The coherent state $\alpha$ is the quantum
mechanical counterpart of monochromatic light. It can be represented
by amplitude and phase in a manner $\alpha=\left|\alpha\right|e^{i\theta}$. The complex number specifies the amplitude in photon number units.

\subsection{Photon statistics within a time segment follows a Poisson
distribution.}

We know that a perfectly coherent light follows Poisson distribution
\cite{Mark Fox}. Consider a beam segment corresponding to a predefined
time segment with an average number of photon statistics as $\mu=\phi T$
where, $\phi$ is an average optical flux. We divide the time segment
into small time bins and show that the probability of finding $n$
photons within a time segment $(T)$ containing $N$ time bins follow
Poisson distribution. Let this probability be given by $P(n)$.
Consider probability of finding $n$ time bins containing $1$ photon
is $p_{1}=\mu/N$ and $N-n$ time bins containing no photons. This
probability is given by a binomial distribution
\begin{equation}
\begin{array}{ccc}
P(n) & = & \frac{N!}{n!(N-n)!}p_{1}^{n}(1-p_{1})^{N-n}\\
 & = & \frac{N!}{n!(N-n)!N^{n}}\mu^{n}\left(1-\frac{\mu}{N}\right)^{N-n}
\end{array}\label{eq:photon statistics}
\end{equation}
We will  take the limit as $N\rightarrow\infty$, the
probability is $\underset{N\rightarrow\infty}{\textrm{lim}}[P(n)]=\frac{1}{n!}\mu^{n}e^{-\mu}$,
thus, the probability of finding $n$ photons in $T$ time segment
follows Poisson distribution $P(n)$. It is important to mention that we have considered ideal detectors with $100\,\%$ efficiency.

\subsection{Quantum theory of photon detection}

When conducting an experiment to leverage the Poisson nature of photon
statistics of coherent light we have to consider the optical loss
and imperfection in the devices. These are inefficient optics, absorption
and imperfect detectors. These lead to a random sampling of photons
which degrades the photon statistics. If we look at the quantum theory
of photon detection \cite{Mark Fox}, it aims to connect the photo count statistics
of the detector within a time segment $T$ with photon statistics
impinging at the detector. The variance in the photo count number is
$\left(\triangle C\right)^{2}$ and the variance in the photon number
is $\left(\triangle n\right)^{2}$. The relationship is established
by
\begin{equation}
\left(\triangle C\right)^{2}=\eta^{2}\left(\triangle n\right)^{2}+\eta(1-\eta)\mu.\label{eq:photon statistics}
\end{equation}
If the detector was perfect (i.e. $\eta=1)$ then the photon
count statistics would have been equal to the photon statistics. Consider
a coherent source (i.e. $\left(\triangle n\right)^{2}=\mu$) and an imperfect detector as in case of PMT or SPAD, the equation \ref{eq:photon statistics}
will become $\left(\triangle C\right)^{2}=\mu\eta=C$ thus, the photo count
statistics $(C)$ and photon statistics $(\mu)$ both follow Poisson
distribution for all values of detection efficiencies.

\section{Time of arrival generators\label{sec:Time-of-arrival}}

The time of arrival based QRNG systems are based on encoding the arrival time
of photons or pulses. During the short time periods, the arrival of a photon at the detector
follows an exponentially distributed time $\lambda e^{-\mu T}$. The
time between the two arrivals is the difference between two exponential
random variables which is also exponential. The randomness in the
exponential distribution is converted to a uniform bit sequence using
post-processing algorithms. Another way to flatten the exponential
distribution is by taking short time bins from an external reference
and considering the time of arrivals within those bins. Nie \textit{et
al.} \cite{Nie} has explained that when randomness is extracted
from the mere arrival of time then the generated random numbers are biased.
They have proposed a new method to generate uniform random numbers
from the photon clicks at a fixed time duration $(t,t+T)$. The fixed time period is divided into small time bins with precision $t_{min}$. The time period is always less than the dead time allowing single detection. The advantage of this method is twofold, low-bias and large throughput compared to other methods of quantum
random number generation.

\subsection{Theoretical Analysis}

The photon flux which is an average number of photons passing through
a cross-section of a coherent beam follows a Poisson process. Precisely,
a coherent beam with well-defined average photon number will follow
a photon number fluctuation at a short time interval. The reason behind
this fluctuation is due to the fact that we cannot predict the position
of these photons. Consider a 1550 nm laser emitting 0 dBm of power,
this will have an average flux of $7.78\times10^{15}\textrm{ photons }s^{-1}$.
If we apply $60$ dB attenuation the average flux will be $7.7\times10^{9}\textrm{ photons }s^{-1}$.
We can interpret this as an average of $7.7$ photons in the 1 ns time
segment. Let us consider a time segment of 100 ps corresponding to
an average flux of $0.77$. To summarize, if the coherent beam is attenuated
to an extent that the beam segment contains few photons say on an
average of $0.77$ photons and we make measurements of some $n$ samples,
then, one can observe the random fluctuations in the photon number.
This comes from the fact that the stimulated emission in the semiconductor
laser is inherently random. Consider a time segment $T$, with mean
photon number of $\mu T$, the probability distribution of $k$ photons
arriving in time interval $T$ is given by $P(k)=\frac{e^{-\lambda\textrm{T}}(\mu T)^{k}}{k!}$.
The photon number follows a Poisson distribution. Hence, the time
interval between the arrival of consecutive photons also follows the Poisson
process \cite{Wayne}.
\begin{figure}[b]
\begin{centering}
\includegraphics[scale=0.4]{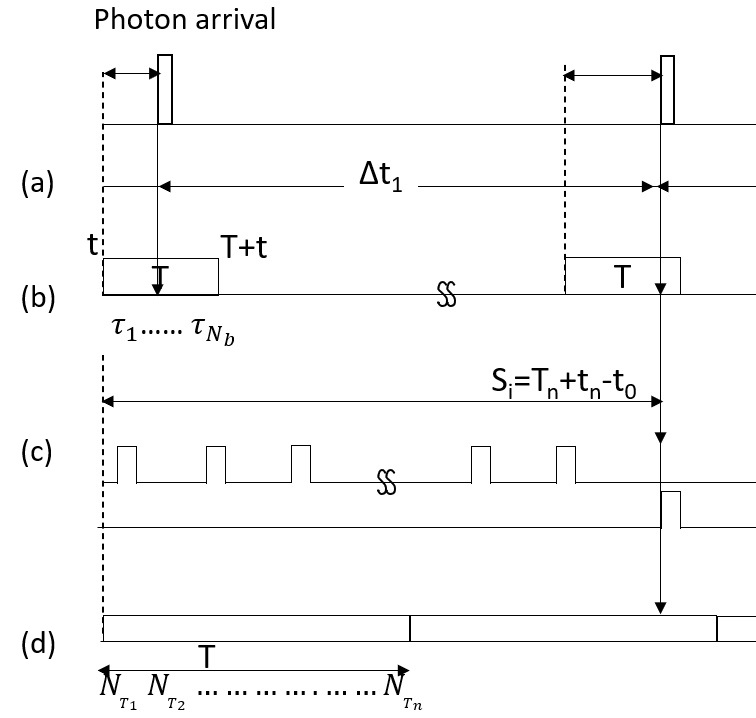}
\caption{Timing diagram for arrival time of photon using different methods
a) Wayne \textit{et al.} \cite{Wayne}, b) Nie \textit{et al.} \cite{Nie}
, c) Yan \textit{et al.} \cite{Yan} and d) method proposed. \label{fig: method of TOA}}
\end{centering}
\end{figure}
Consider $\lambda T=0.1$, there is $90.5\%$ probability of detecting no photons in $T$ time segment, $9\%$ of detecting single photons and finite probability of $0.5\%$ of multi photon in $T$ time segment. The time period is divided into
$N_{b}$   time bins and each time bin is $\tau_{i}=(\frac{i-1}{N_{b}}T,\frac{i}{N_{b}}T)$.
For an ideal detector  $(\eta=1)$,  there would have been multiple clicks in $T$, and first
detection will be the minimum value of the random variable. However, the dead time is more than $T$ as a result there is a single detection in $T$. For a
detection event the conditional probability of getting a detection
at $i^{th}$ position given $k$ photons appear in that period is given
by $P(\hat{n}=i\left|k\right)$. It is probability that detection
happened at $\tau_{i}$ when $k$ photons are present in a period.

\[
\begin{array}{ccc}
P(\hat{n}=i\left|k\right) & = & \frac{P(i,k)P(T-\tau,k=0)}{P(k)}\\
 & = & \frac{\frac{e^{-\lambda\tau_{i})}(\lambda(\frac{T-(i-1)T}{N_{b}}-\frac{T-iT}{N_{b}}))^{k}}{k!}}{\frac{e^{-\lambda T}(\lambda T)^{k}}{k!}}e^{-\lambda(T-\tau_{i})}\\
 & = & \left(\frac{1-(i-1)}{N_{b}}-\frac{1-i}{N_{b}}\right)^{k}\\
 & = & \left(\frac{1-(i-1)}{N_{b}}\right)^{k}-\left(\frac{1-i}{N_{b}}\right)^{k}
\end{array}
\]

The probability distribution function  of the arrival of a photon
conditioned on the fact that only $1$ photon is available in the time period
$T$ is

\begin{equation}
\begin{array}{ccccccc}
P(i,k=1) & = & \frac{1}{N_{b}} & = & \frac{1}{T/\tau} & = & \frac{\tau}{T}\end{array}.
\end{equation}

Given a time period \cite{Nie,Yan},  the probability of a photon
to arrive at each time bin is $\frac{1}{N_{b}}$. In this expression,
$\tau$ is the independent variable of the probability distribution function.
The probability density is given by $\frac{1}{T}$. Thus, the arrival
time is uniformly distributed in $[0,T].$ We have used binary code
to encode the time bins. In Fig. \ref{fig: method of TOA}, we have
presented  different methods for implementing time of arrival based QRNG.
Wayne \textit{et al.} \cite{Wayne} extracted random numbers by translating time intervals between detection
into time bins. Nie \textit{et al.} \cite{Nie} generated  raw quantum random numbers by considering time difference between photon click and an external time reference. The distribution of time difference between the external reference
clock and photon click is approximately uniform. Yan \textit{et al.}
\cite{Yan} have generated the highest reported raw data bits of $128$
Mbps  by measuring the time of arrival from a common starting point. They convert each arrival time into sum of fixed period and phase time. Thereafter, they generate random numbers from phase time.  In this work we address the arrival time of photon differently, we have considered
an external reference as the basis of generation of raw bits. We have
divided the external time reference $T$ into $N_{T}$ divisions.
The arrival time is given by
\begin{equation}
A_{n}=\textrm{mod}[N_{b},N_{T_{i}}]
\end{equation}
where, $N_{b}$ is the total number of random digits we want to generate. We do not restrict ourselves to modular arithmetic, the purpose is to segregate the time segment $T$ into $N_{T}$ fragments with each fragment equal to the precision of the measurement device. These fragments are distributed into $N_{b}$ divisions to generate $N_{b}$ random digits.
The arrival of photon will be randomly falling in the range
$1\leq N_{T_{i}}\leq N_{T}$, hence, applying the above equation \cite{Wayne,Nie} we can prove that $N_{T_{i}}$ is uniformly distributed
in $[0,T)$.  In Table \ref{tab:comparison},
we have compared the proposed work with existing works.

\begin{table}[b]
\caption{Comparison of proposed QRNG with existing works on the basis of nature of source, detector, external reference clock, entropy per detection, precision of time measurement and throughput. CS stands for coherent
source. \label{tab:comparison}}
\begin{ruledtabular}
\begin{tabular}{ccccccccc}

Ref. & Source & Detector &  Ext ref  &  Entropy &  Resolution & Rate\tabularnewline
 &  &  &   (ns) & (ns) &  &  (Mbps)\tabularnewline
\colrule
\cite{Wayne} & LED & SiAPD &  - &  5.5 & 5 & 40 \tabularnewline

\cite{Nie} & CS & SiAPD &  40.96  &  8 & 0.160 & 109 \tabularnewline

\cite{Yan} & CS & SiAPD &  20  &  8 & 20 & 128 \tabularnewline

AB & CS & InGaAs & 500  &  16 & 0.005 & 2.4 \tabularnewline

\end{tabular}
\end{ruledtabular}
\end{table}

\subsection{Source of bias }

To quantitatively evaluate the randomness of the raw data, we need
to model the system carefully and figure out the facts that would
introduce bias. There are a few major device imperfections to be examined.
\begin{enumerate}
\item The laser intensity needs to be constant. We have plotted the number
of detections per $101 \mu s$ to validate that the average photon
count statistics is uniform. If the average photon number is 0.1 for
100 ns  and $10\,\mu s$ dead time of detector. There should
be one detection every 10100 ns  or 10 clicks in
101 $\,\mu s$. We take 100 samples of 1010$\,\mu s$ intervals, and
validate the mean photon number.

\item Detector dark counts are random clicks in absence of photons. It will be  interesting to analyze the effect of  random noise from dark counts getting mixed with  random numbers generated from arrival times of photons.   In the experiment, the dark
count rate is roughly $350-400$ cps. Compared to the detection count rate,  the proportion of dark count is much less therefore, it is not possible to conclusively state whether it  introduces bias or not.
\item In one of our implementations, we have considered a dead time of $5\,\mu s$,
in other, we have considered a dead time of $10\,\mu s$,
which is far greater than the duration of external reference which
is 100 ns (500 ns). The dead time can be considered as a drift \cite{Nie}
and it does not affect the quality of random numbers.
\item The probability for multi-photon emission from an attenuated CW laser
is non-zero. If we consider detectors that can distinguish between
multi-photon and single-photon then we can discard the multi-photon
cases. This will reduce the bias in the output. However, we have
considered the mean photon number much less than 1 and this would reduce
the chances of multi-photon considerably.
\end{enumerate}

\section{Experimental analysis\label{sec:Experimental-analysis}}

\begin{figure}[b]

\includegraphics[scale=0.55]{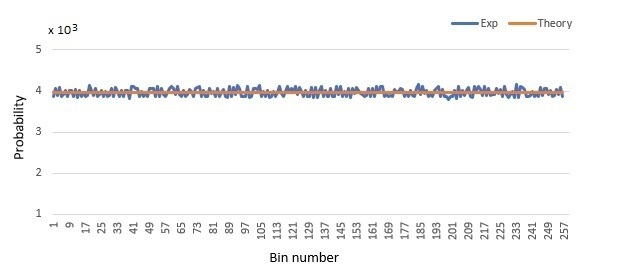}

\caption{Probability distribution for theoretical vs experimental for
256 time bins with detection event of 90 Kb raw data.\label{fig:Probability-distribution-for}}

\end{figure}

The experimental setup is presented in Fig.  \ref{fig:Experimental-setup},
it comprises of Distributed Feedback (DFB) Laser operated in continuous
mode. The laser diode has a wavelength of $1550$ nm and output power
of $0.1$ mW. We have $2$ variable optical attenuators to tune the amplitude
of weak coherent source to the desired value. One of them is kept fixed
and the other is altered to achieve granularity. We have implemented  SPD from
CHAMPION Aurea detectors in free-running mode. It has the flexibility of adjusting at variable efficiencies, for example, $10\%$, $20\%$, and $30\%$  and the dead time can be configured
to achieve the required count rate with respect to particular efficiency.
We have considered different values of $N_{b}$ as 8, 100, 200, 256, 500 and 512 for external clock of 100 ns.  When we implemented the scheme with  500 ns clock reference,  we have considered 65536 divisions and generated 16 bits per detection. This implies
that each division is 7.6 ps. The jitter in SPD is 180 ps and hence
for better results, we should consider a division greater than this
period. At $10\%$ efficiency and 10 $\mu s$ dead time the SPD has a 350-400 dark counts
and counting rate of 90 Kcps, the photon flux at the input of SPD
is $9\times10^{6}$ cps corresponding to -89 dBm optical power with
mean photo count of 0.09. When SPD has a counting rate of 96 Kcps,
the photon flux at the input of SPD is $2.4\times10^{7}$cps corresponding
to -85 dBm optical power with a mean photo count of 0.24.
However, at a continuous rate, we have considered the SPD count rate as $9\times10^{4}$.
We have presented the frequency distribution of the digits 1 to 256
generated randomly in Fig. \ref{fig:Probability-distribution-for}.
We find that the throughput is almost uniform. We have converted this
to binary and tested it on the NIST test platform. We have performed Toeplitz hashing,
however, the results were not improving the ENT or  NIST tests. The raw numbers generated are unpredictable
and uniform simultaneously, hence eliminating the need for applying any mathematical algorithm. At 5 $\mu s$ dead time we have considered
a counting rate of 150 Kcps. We have used single time-to-digital
converter (TDC), TDC\_AS6501 in field-programmable gate array (FPGA)
which is Zynq UltraScale+ MPSoC for post-processing the data. The
TDC can respond to external clock reference from $2$ MHz to $12.5$ MHz. The TDC has an internal clock of 5 ps and it
can count till 500 ns. An external reference with frequency 10 MHz in one implementation and 2 MHz in other implementation with a jitter of  3 ps is used as a reference clock which is edge synchronized with the TDC counter. The throughput at 150 Kcps count rate is 2.4 Mbps as raw QRNG data with $5\mu s$ dead time.

\begin{widetext}
\begin{center}
\begin{figure}[b]

\includegraphics[scale=0.35]{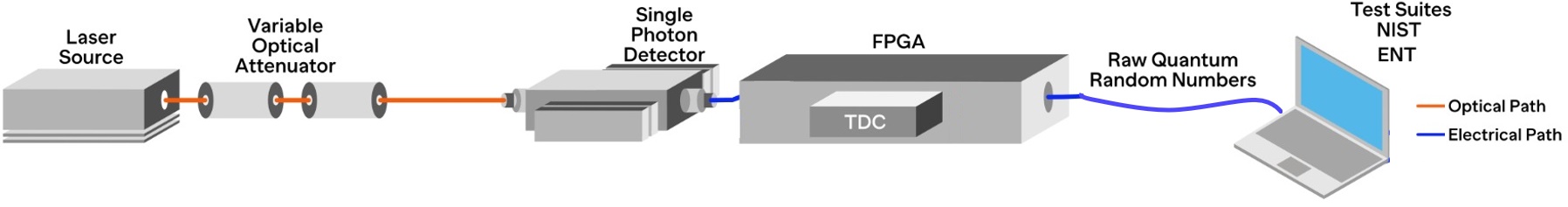}

\caption{Experimental setup. \label{fig:Experimental-setup}}

\end{figure}
\end{center}
\end{widetext}

\begin{table*}
\caption{Results of ENT tests for a typical sequence of 1069712 bits.\label{tab:Results-of-ENT}}
\begin{ruledtabular}
\begin{tabular}{llll}

ENT test item & QRNG & TRNG & Ideal value\\
\colrule

Entropy (bits per bit) & 1.000000 & 1.000000 & 1.000000\\
Chi-square distribution & 45\% & 15.11\% & 10\% $\sim$ 90\%\\
Arithmetic mean value & 0.4997 & 0.5001 & 0.5000\\
Monte Carlo value for Pi & 3.1515369080 & 3.142180307 & 3.1415926536\\
Serial correlation coefficient & 0.000590 & 0.000088 & 0.000000\\

\end{tabular}
\end{ruledtabular}
\end{table*}

\section{Entropy estimation\label{sec:Post-Processing}}

Entropy in the information-theoretic sense is a measure of randomness
or unpredictability of the outputs of an entropy source. The larger the entropy, the greater the uncertainty in predicting
the outcomes. Estimating the amount of entropy available from a source
is necessary to see how many bits of randomness are available. If
a discrete random variable $X$ has $n$ possible values where, the
$i^{th}$ outcome has probability $p_{i}$ then, the Rényi entropy
of order $\alpha$ is defined as

\begin{equation}
H_{\alpha}(X)=\frac{1}{1-\alpha}\log_{2}\left(\stackrel[i=1]{n}{\sum}p_{i}^{\alpha}\right)
\end{equation}

for $0\,\le\,\alpha\,\le\,\infty$. As $\text{\ensuremath{\alpha\rightarrow\infty}}$,
the Rényi entropy of $X$ converges to the negative logarithm of the
probability of the most likely outcome, called the min-entropy

\begin{equation}
H_{\infty}(X)=\underset{\alpha\rightarrow\infty}{\lim}H_{\alpha}(X)=-\log_{2}\max p_{i}.
\end{equation}

The name min-entropy $(H_{\infty})$ stems from the fact that it is the
smallest in the family of Rényi entropies. In this sense, it is the most
conservative approach of measuring the unpredictability of a set of outcomes to measure the randomness content of a distribution.
The standard Shannon entropy (which measures the average unpredictability
of the outcomes) offers only a rough estimation of randomness. On
the other hand, $H_{\infty}$ is used as a worst-case measure of the
uncertainty associated with observations of $X$. This represents
the best-case work for an adversary who is trying to guess an output
from the noise source. For $H_{\infty}$, $p_{i}$ is the detection event probability at $i^{th}$ time bin and
\begin{equation}
 H_{\infty}=-\frac{\log_{2}P_{max}}{\log_{2}N_{b}}=-\frac{\log_{2}P_{max}}{\textrm{bit}}=0.9971 \end{equation}
 from the maximum frequency of 0.00397 which is higher than \cite{Nie}.
In Fig. \ref{fig:Probability-distribution-for}, the frequency distribution is almost uniform, the experimental value is  close to theoretical value, in that case the entropy is approximately equal  to 1 with high precision hence, in some works   \cite{Yan}, entropy extraction is addressed by Shannon entropy.
\begin{equation}
H=-\stackrel[i=1]{N_{b}}{\sum}p_{i}\textrm{log\ensuremath{_{2}}}p_{i},
\end{equation}
The probability of detection event at each time bin is uniform
thus, $p_{i}=\frac{1}{N_{b}}$. Hence, $H=\textrm{log\ensuremath{_{2}}}N_{b}$,
which means if we divide the time segment into $N_{b}=8, 256$ or 1024
bins, we can generate 3, 8, or 10 random bits per photon arrival.

We have generated a 35 Gb random bitstream. Precisely, 63 files of 80 Mb
data (total 5 Gb data) leading to raw QRNG data of 35-45 Gb. We performed
30 runs of 180 Mb random data files in the NIST STS test suite \cite{NIST}. Raw QRNG data
is of length 186504193 bits. We have tested 100 sequences of sample size $10^{6}$ bits. Some important
statistical properties of generated random bitstreams are computed
using the ENT program \cite{ENT}. ENT is a series of basic statistical
tests that evaluate the random sequence in some elementary features
such as the equal probabilities of ones and zeros, the serial correlation,
etc. The testing results with ENT are presented in Table \ref{tab:Results-of-ENT}.
Each test in the NIST suite evaluates a p-value which should be larger than the
significance level. The significance level in the tests is $\alpha=0.01$.
The test is considered successful if all the p-values satisfy 0.01
\ensuremath{\le} p-value \ensuremath{\le} 0.99. Ten random data files
with each file size of 100000 bits are tested. In the tests producing
multiple outcomes of p-values, the worst outcomes are selected. The
testing results with NIST are presented in Table \ref{tab:Results-of-NIST}.
All the output p-values are larger than 0.01 and smaller than 0.99,
which indicates that generated random bits well pass the NIST tests.
We have compared the raw quantum random numbers with the NIST's Blum-Blum-Shub
algorithm and in-house TRNG built from FPGA from the asynchronous sampling
of a ring oscillator.

\begin{table}[H]
\caption{Results of NIST tests for a typical sequence of 100000 random bits.\label{tab:Results-of-NIST}}
\begin{ruledtabular}
\begin{tabular}{llll}
NIST test item & BBS & QRNG & TRNG\\
 & p value & p value & p-value\\
\colrule
Frequency  & 0.816537 & 0.534146 & 0.213309 \\
Block frequency & 0.366918 & 0.122325 & 0.015598 \\
Cumulative sums & 0.955835 & 0.634146 & 0.851383 \\
Runs & 0.090936 & 0.739918 & 0.137282 \\
Longest run & 0.202268 & 0.534146 & 0.494392 \\
Rank & 0.202268 & 0.534146 & 0.383827 \\
FFT & 0.275709 & 0.350485 &0.494392 \\
Non overlapping template & 0.764295 & 0.991468 & 0.883171 \\
Overlapping template & 0.455937 & 0.350485 & 0.779188 \\
Universal & 0.060806 & 0.213309 & 0.699313 \\
Approximate entropy & 0.971699 & 0.839918 & 0.739918 \\
Random Excursions & 0.534146 & 0.739918 & 0.186566 \\
Random Excursions Variant & 0.911413 & 0.457799 &0.311542 \\
Serial & 0.739918 & 0.911413 & 0.779188 \\
Linear complexity & 0.145326 & 0.739918 & 0.289667 \\
\end{tabular}
\end{ruledtabular}
\end{table}

\section{Conclusion\label{sec:Conclusion}}

We have designed and tested a practical high-speed QRNG based on the
 time of arrival quantum entropy from a CW laser at telecommunication wavelength.
This is the first work on time of arrival QRNG using InGaAs detectors. These detectors
have a greater dead time than Silicon detectors enabling us to increase
the external reference time to 500 ns compared to previous values
of 40.6 ns \cite{Nie} and 20 ns  \cite{Yan}. We have implemented precision time measurement of 5 ps which is reported for the first time. Hence, we could extract 16 bits of entropy from one photon
arrival time. The photon arrival follows a Poisson distribution, an exponential waiting time introduces bias, this is overcome
using external reference clock and we have generated uniform random
numbers with high entropy particularly, min-entropy always greater
than 0.99 value. The method of time measurement of photon is simpler in implementation and  higher in precision time measurement than earlier works \cite{Nie,Yan,Wayne}. The proposed work can also be used
to generate higher throughput by increasing the duration of the external reference
clock however, the increase in throughput will be linear.  We propose that implementing the time of arrival QRNG with InGaAs detectors and high precision time measurement
will enable  generating maximum entropy per detection event.
The proposed work  eliminates the use of any mathematical algorithm
to generate uniform output hence, the random numbers produced from
the system are derived from the quantum behavior of photons.

\begin{acknowledgments}

We are thankful to  M.T. Karunakaran for bringing deep technical insights into the project.
\end{acknowledgments}

%

\end{document}